\documentclass{article}

\usepackage{microtype}
\usepackage{graphicx}
\usepackage{booktabs} %

\usepackage{hyperref}

\usepackage[accepted]{icml2025}

\usepackage{amsmath}
\usepackage{amssymb}
\usepackage{mathtools}
\usepackage{amsthm}
\usepackage{enumitem}
\usepackage{caption}
\usepackage{subcaption}

\usepackage[capitalize,noabbrev]{cleveref}

\theoremstyle{plain}

\theoremstyle{definition}

\theoremstyle{remark}

\usepackage[textsize=tiny]{todonotes}

\usepackage{titlesec}
\titlespacing*{\section}
{0pt}{2pt plus 0pt minus 0pt}{0pt plus 0pt}
\titlespacing*{\subsection}
{0pt}{1pt plus 0pt minus 0pt}{0pt plus 0pt}

\newcommand{\topic}[1]{\noindent\textbf{#1}}

\newcommand{\ourtitle}{%
    MADCAT: Combating Malware Detection Under Concept Drift \\ with Test-Time Adaptation
}

\icmltitlerunning{MADCAT}

\begin{document}

\twocolumn[
\icmltitle{\ourtitle}

\icmlsetsymbol{equal}{*}

\begin{icmlauthorlist}
\icmlauthor{Eunjin Roh}{osu}
\icmlauthor{Yigitcan Kaya}{ucsb}
\icmlauthor{Christopher Kruegel}{ucsb}
\icmlauthor{Giovanni Vigna}{ucsb}
\icmlauthor{Sanghyun Hong}{osu}
\end{icmlauthorlist}

\icmlaffiliation{osu}{Oregon State University, Corvallis, OR USA}
\icmlaffiliation{ucsb}{University of California, Santa Barbara, CA USA}

\icmlcorrespondingauthor{Eunjin Roh}{rohe@oregonstate.edu}

\icmlkeywords{Machine Learning, ICML}

\vskip 0.3in
]

\begin{abstract}
We present MADCAT%
\footnote{\textbf{MA}lware \textbf{D}etection under \textbf{C}oncept Drift through \textbf{A}daptation during \textbf{T}est-Time}, a self-supervised approach designed to address the concept drift problem in malware detection. MADCAT employs an encoder-decoder architecture and works by test-time training of the encoder on a small, balanced subset of the test-time data using a self-supervised objective. During test-time training, the model learns features that are useful for detecting both previously seen (old) data and newly arriving samples. We demonstrate the effectiveness of MADCAT in continuous Android malware detection settings. MADCAT consistently outperforms baseline methods in detection performance at test-time. We also show the synergy between MADCAT and prior approaches in addressing concept drift in malware detection.
\end{abstract}

\section{Introduction}
\label{sec:intro}

Malware is \emph{any} program that, when downloaded and executed on a victim's machine, exhibits malicious behavior---such as privilege escalation, data exfiltration, or backdoor injection~\cite{sikorski2012practical, %
king2021data, giorgio2021backdoor}. In recent years, the volume of malware threats has increased at an unprecedented scale, far exceeding what analysts and security experts can address manually. As a result, machine learning (ML) has emerged as a promising and scalable solution for combating this ever-evolving threat landscape~\cite{sahin2021a, mohamad2021static, chaulagain2020hybrid, yuan2020bytelevel, nataraj2011malware, arp2014drebin}. ML-based malware detectors are trained on large collections of malicious and benign samples, and then deployed in the wild to automatically determine whether a suspicious program is malware or not.

\begin{figure}[t]
    \centering
    \includegraphics[width=\linewidth]{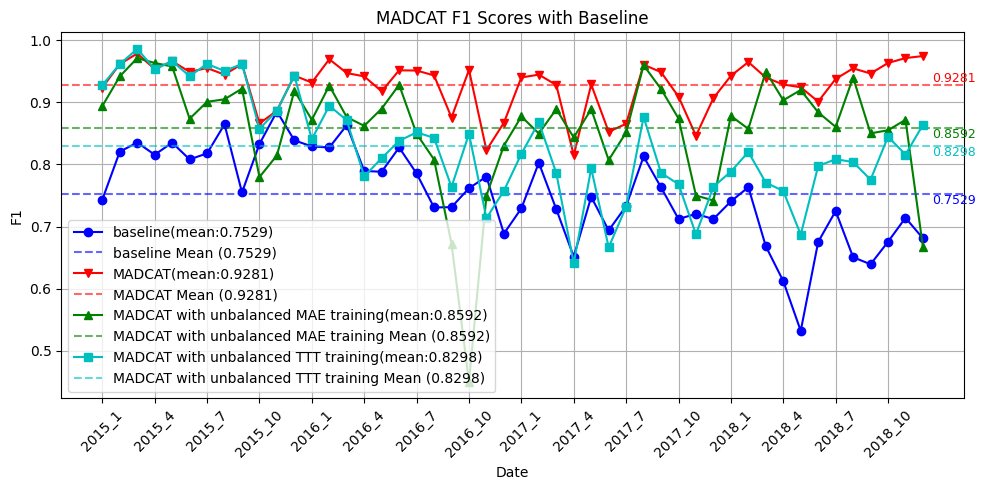}
    \vspace{-2.0em}
    \caption{\textbf{MADCAT Performance.} 
    MADCAT consistently outperforms the baselines across all evaluation periods, addressing the concept drift in ML-based malware detection.}
    \label{fig:comparison}
    \vspace{-1.0em}
\end{figure}

However, this emerging paradigm has faced resistance due to a real-world challenge: \emph{concept drift}~\cite{schlimmer1986beyond}---the phenomenon where detectors, once deployed in the wild, gradually become ineffective as the characteristics of malware
evolve over time~\cite{chen2023continuous, chen2023is}. Most prior work addresses this issue with \emph{supervised learning} techniques that allow models to learn \emph{new} features useful for detecting emerging malware variants. This is typically achieved through re-training or fine-tuning detection models on newly collected and labeled datasets~\cite{monlinacoronado2023efficient, chen2023continuous, li2025revisiting}. While shown to be effective, these approaches rely on the availability of high-quality labels for incoming future data---a requirement that is often difficult to satisfy in real-world scenarios~\cite{wu2023from, zhu2020measuring, joyce2023motif}.

\topic{Contributions.}
In this paper, we address the concept drift problem through an orthogonal approach: \emph{self-supervised learning}. Instead of relying on detection models to extract useful features from high-quality labeled data, our method reduces the dependency on such supervision. Instead, it focuses on learning robust representations that preserve generalization performance over time~\cite{hendrycks2019using}, even as malware evolves. Importantly, our approach is complementary to supervised learning techniques---meaning it can be combined with them to achieve synergistic benefits when a small set of labeled future data is available.

\begin{figure*}[t]
    \centering
    \includegraphics[width=0.7\linewidth]{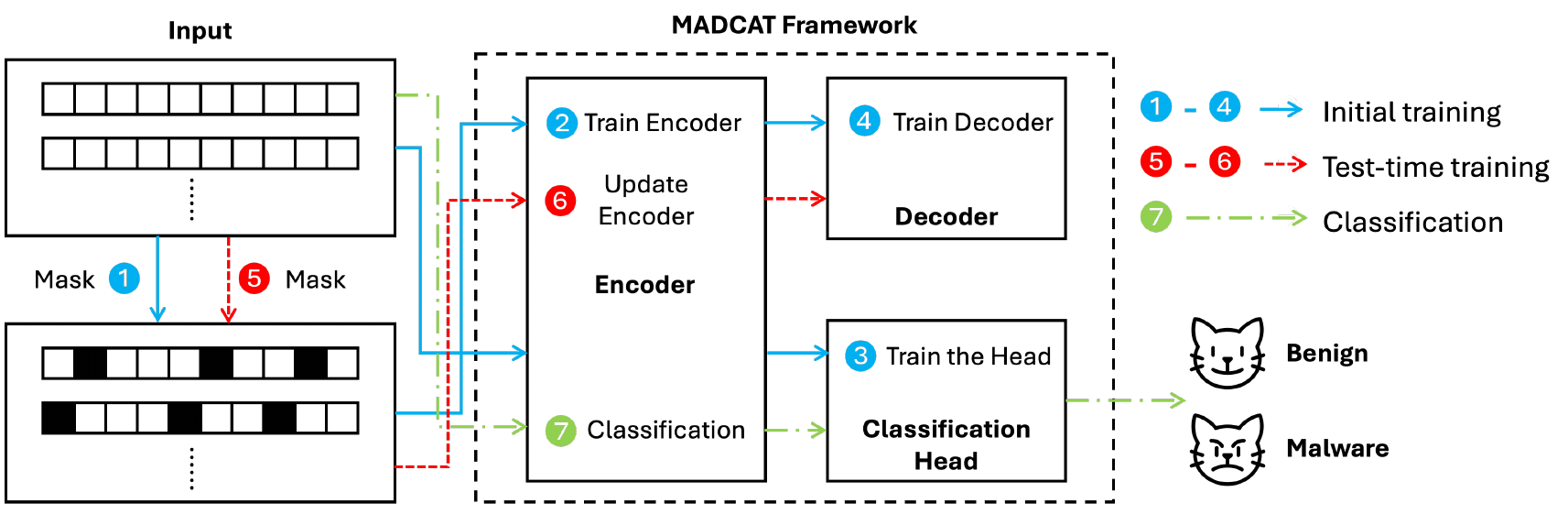}
    \vspace{-0.8em}
    \caption{\textbf{MADCAT Workflow.} 
    }
    \label{fig:overview}
    \vspace{-1.0em}
\end{figure*}

To demonstrate the potential, we present MADCAT, a self-supervised malware detector that performs test-time adaptation on future malware data using the masked autoencoder (MAE) approach~\cite{gandelsman2022testtime}. The MAE is first trained to reconstruct samples in which a random subset of the input has been masked. This form of self-supervision encourages the encoder to learn representations (or features) of both malware and benign samples that are robust to distributional shifts over time. At test-time, MADCAT performs fine-tuning of its encoder on future data.

We evaluate MADCAT on %
Android malware detection using a dataset consisting of malware and benign applications collected over a 7-year period. %
As summarized in Figure~\ref{fig:comparison}, our method maintains the performance over time and shows a better F1 score compared to the baseline approaches that rely on supervised learning.
Our evaluation also highlights the MADCAT configurations that yield the best performance, such as balancing the test-time data used for fine-tuning the autoencoder. Moreover, we show that supervised learning can serve as a strong complement to our approach, showing performance synergy when combined with MADCAT. We hope our work draws the attention of the test-time adaptation community to malware detection and inspires future research that integrates self-supervision.

\section{Background and Related Work}
\label{sec:prelim}

\topic{Concept drift}
refers to the gradual change in the statistical properties of data over time. It was first introduced by~\citet{schlimmer1986beyond}, and has since been further studied in subsequent works aimed at identifying and mitigating its effects~\cite{tsymbal2004the, gama2014a, morenotorres2012a, jordaney2017transcend, barbero2022transcending, chen2023is, chen2023continuous, li2025revisiting}.
In malware detection, concept drift can significantly degrade model performance as their static structure---such as their functions and components---changes over time.

There has been progress in addressing concept drift in malware detection~\cite{jordaney2017transcend, barbero2022transcending}. 
The most recent works employ \emph{supervised learning}.
\citet{chen2023continuous} adopt active learning. They encourage the encoder to map similar samples to nearby embeddings and iteratively expand the training set by selecting the most uncertain samples. The detector is then retrained on this expanded dataset using ground-truth labels. \citet{chen2023is} identified two types of concept drift---feature-space drift and data-space drift---and found that data-space drift has a substantial impact on detection performance. To address this, they employ online learning based on pseudo-labels.
In contrast, MADCAT employs a self-supervised strategy that operates using pseudo-labels.

\topic{Test-time adaptation}
is the process of adjusting a pre-trained model to unlabeled target domain data at inference time. It has emerged as a promising solution to mitigate performance degradation caused by distributional shifts when the traditional model only focuses on a limited distribution for both training and testing~\cite{liang2025a}.

\emph{Test-time training} is a form of test-time adaptation that fine-tunes a model during inference. The concept dates back to~\citet{bottou1992local}, but its first application to modern computer vision tasks was introduced by~\citet{sun2020testtime}, who proposed a self-supervised approach that updates the model using a single unlabeled test sample. Since then, several works have advanced test-time training techniques across visual and textual domains~\cite{gandelsman2022testtime, wang2023testtime, hardt2024testtime}. However, few studies have explored test-time adaptation for addressing concept drift in malware detection.

\citet{alam2024morph} is the most recent work to adopt the concept of test-time training for addressing concept drift in malware detection. They perform pseudo-labeling of unlabeled test samples and fine-tune the classifier using high-confidence pseudo-labeled data. However, as shown in \S\ref{subsec:synergy}, this approach remains ineffective under long-term concept drift (over a period of $\sim$2 years), and notably, the concept of self-supervision has not been leveraged. 

MADCAT shows the effectiveness of using self-supervision in malware detection for addressing concept drift, while also forming a synergistic combination with pseudo-labeling.

\section{MADCAT}
\label{sec:method}

Figure~\ref{fig:overview} shows the overall workflow of MADCAT.
The detector has two components:
MAE for self-supervised test-time training and a classification head for malware detection.

\subsection{Initial (Training-time) Training}
\label{subsec:initial}

MADCAT first performs an initial training.
During this phase, a portion of the input features is randomly masked based on a predefined masking ratio (0.0--0.9).
The encoder is trained to capture meaningful representations from the partially masked input, while the decoder learns to reconstruct the masked features.
Once the MAE is trained, the encoder is frozen.
A separate classification head (detector) is then trained on the unmasked input data, using the representations produced by the frozen encoder.

\subsection{Test-time Adaptation with Self-supervision}
\label{subsec:self-supervision}

Once deployed, MADCAT performs test-time adaptation to combat concept drift. 
Because obtaining high-quality, human-annotated labels for newly emerging malware at test time is often impractical~\cite{wu2023from, zhu2020measuring, joyce2023motif}, we adopt a self-supervised test-time training approach proposed by~\citet{he2022masked}.
At test time, each new input sample is partially and randomly masked and passed through the MAE. 
The encoder is then fine-tuned by minimizing the reconstruction loss. 
After this adaptation step, the updated encoder is paired with the pre-trained classification head to perform malware detection.
This approach enables MADCAT to adapt to distributional shifts in the data 
without requiring labeled test-time samples.

\subsection{Handling Class Imbalance with Pseudo-Labeling}
\label{subsec:label-requirements}

Prior work used highly imbalanced datasets,
with benign samples significantly outnumbering malware~\cite{pendlebury2019tesseract}.
This can lead the MAE to learn representations of benign data, potentially limiting its ability to generalize well to malware over time.
To address this class imbalance, we incorporate pseudo-labeling into MADCAT. 
These pseudo-labels are generated by the base detection model, 
and the encoder is updated during test-time on a rebalanced dataset prior to classification.
This strategy helps the model maintain strong performance even under skewed data distributions, 
without relying on any human-provided labels.

\section{Evaluation}
\label{sec:results}

\subsection{Experimental Setup}
\label{subsec:setup}

\topic{Datasets.}
We use APIGraph~\cite{zhang2020enhancing}, which contains Drebin~\cite{arp2014drebin} features extracted from Android APKs collected during 7 years (2012--2018).
Data from 2012--2014 is used for the initial training of the MAE and the classification head. 
We split the dataset into 80\% for training and 20\% for validation.

For test-time training, we use the data collected from 2015--2018. To analyze the performance over time, we divided this data by month. Each monthly dataset was further split into 70\% for test-time training and 30\% for validation. These splits are separate from the initial training set and relatively small due to the fine-grained temporal division.

Since the benign dataset is much larger than the malicious dataset (with a ratio of $\sim$9:1), we \emph{randomly downsampled} the benign data to match the number of malicious ones.

\topic{Models.}
We utilize BinaryMLP, a classification model designed for malware detection in recent work~\cite{chen2023continuous}.
We integrate the MAE~\cite{gandelsman2022testtime} into BinaryMLP.
For training, we use a learning rate of 0.003 and 800 epochs for the initial training of the MAE-augmented BinaryMLP.
Test-time training is performed for a single step per sample.
We use cross-entropy loss for training both the MAE and the classification head.

\topic{Metrics.}
We employ two evaluation metrics: \emph{F1 score} and \emph{detection accuracy}.
Our primary results are reported using F1 scores, with the mean F1 score across all monthly datasets provided in parentheses.

\subsection{Effectiveness of MADCAT}
\label{subsec:effectiveness}

\topic{Methodology.}
As a baseline, we trained the BinaryMLP model~\cite{chen2023continuous} on the 2012--2014 data without the MAE module or test-time adaptation.
This baseline model remains fixed after initial training and is not updated thereafter.
The dataset for both initial training of MAE-augmented BinaryMLP and test-time training is balanced based on the ground truth labels.
The default masking ratio of MAE is set to 0.3.
All models are evaluated on the monthly-divided test dataset from 2015--2018.

\begin{figure}[t]
    \centering
    \includegraphics[width=\linewidth]{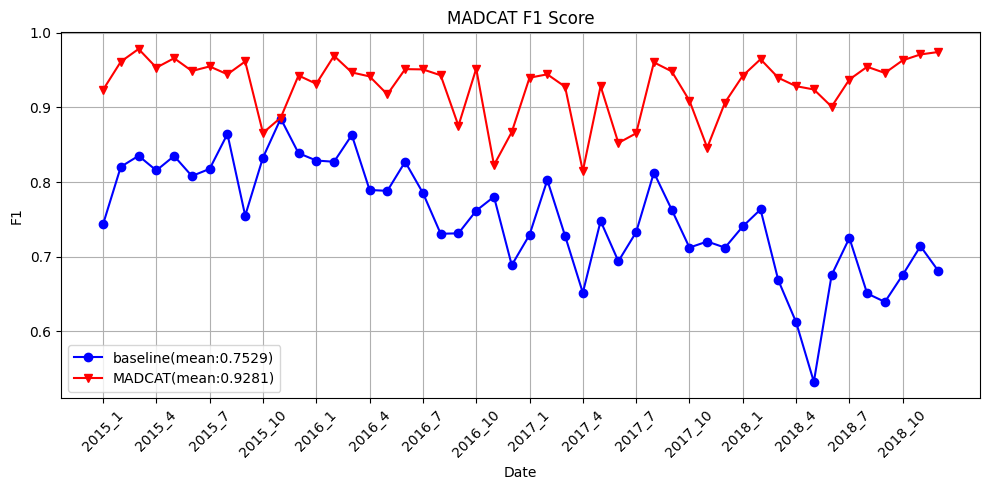}
    \vspace{-2.0em}
    \caption{\textbf{MADCAT Performance.} It maintains consistent F1 scores, while the baseline shows gradual degradation.}
    \label{fig:madcat_result}
    \vspace{-1.0em}
\end{figure}

\topic{Results.}
Figure~\ref{fig:madcat_result} shows the performance of MADCAT over time. 
We report the F1 score for each month, 
and the detection accuracy over the entire period is shown in the legends.
We first show that the baseline model exhibits a gradual decline in performance, indicating the presence of concept drift. In contrast, MADCAT maintains consistent performance, highlighting its robustness to concept drift in the data. Across all cases, MADCAT achieves higher F1 scores than the baseline, demonstrating its effectiveness for malware detection. We analyze the performance on benign and malicious data separately in Appendix~\ref{appendix:madcat_benign_mal}.

\subsection{Synergy with Prior Approaches}
\label{subsec:synergy}

Our previous results assume that ground-truth labels are available at test time to balance the dataset. While it may be feasible for a small subset of the test-time data, obtaining labels for the entire stream of future data is often impractical in real-world scenarios. \citet{alam2024morph} address this issue using pseudo-labeling---where the base model is used to label incoming data, and it is continuously fine-tuned on the newly pseudo-labeled samples. We evaluate whether MADCAT can achieve synergy when combined with pseudo-labeling.

\topic{Methodology.}
Using the same experimental setup as before, we use the BinaryMLP model to generate pseudo-labels for each month's test-time data. However, when the pseudo-labeled data is highly imbalanced (e.g., dominated by one class) MADCAT's performance may degrade due to biased learning~\cite{zheng2022imbalanced}. To address this, we additionally consider three label-balancing strategies:
\begin{itemize}[topsep=0.em, itemsep=0.em, leftmargin=1.4em]
    \item \textbf{Random:} Samples are randomly selected from each pseudo-labeled class to ensure an equal number of benign and malicious samples.
    \item \textbf{Confidence-based top-N:} All samples are sorted by the model's confidence scores within each pseudo-labeled class, and we select the top-N most confident samples.
    \item \textbf{Confidence-based bucket:} We divide the data into 10 buckets based on rounded confidence scores and select an equal number of samples from each bucket to maintain balance across varying confidence labels. %
\end{itemize}

\begin{figure}[t]
    \centering
    \includegraphics[width=\linewidth]{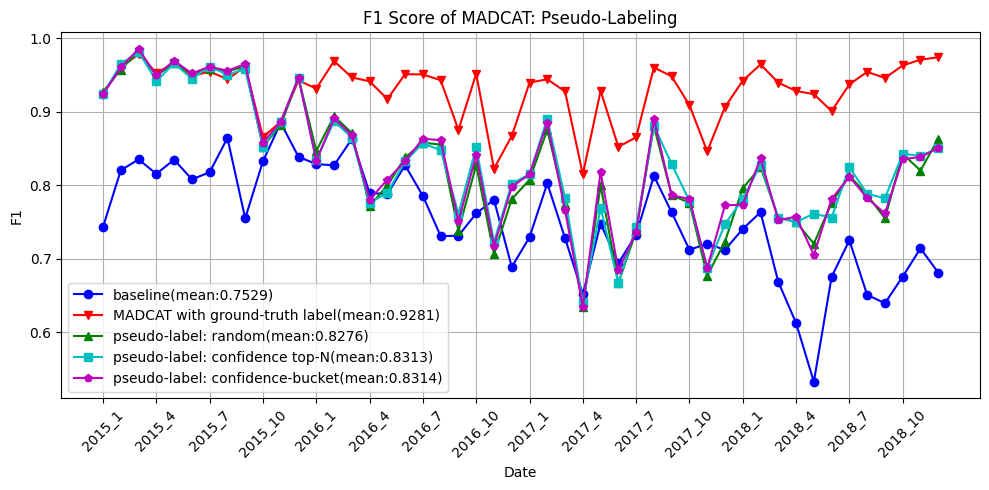}
    \vspace{-2.0em}
    \caption{\textbf{MADCAT performance with pseudo-labeling.}
    All MADCATs achieve higher F1 scores vs. the baseline.}
    \label{fig:madcat_pseudo_label_result}
    \vspace{-1.0em}
\end{figure}

\topic{Results.}
Figure~\ref{fig:madcat_pseudo_label_result} shows our results.
All MADCATs combined with pseudo-labeling achieve higher F1 scores compared to the baseline.
However, the performance is slightly lower than that of MADCAT trained on a dataset balanced using ground-truth labels. This suggest that when the dataset collected for test-time training is unlabeled, pseudo-labeling using the base model can be an effective option.

\subsection{Ablation Study}
\label{subsec:ablation}

\begin{figure}[t]
    \centering
    \begin{subfigure}[h]{\linewidth}
        \centering
        \includegraphics[width=\linewidth]{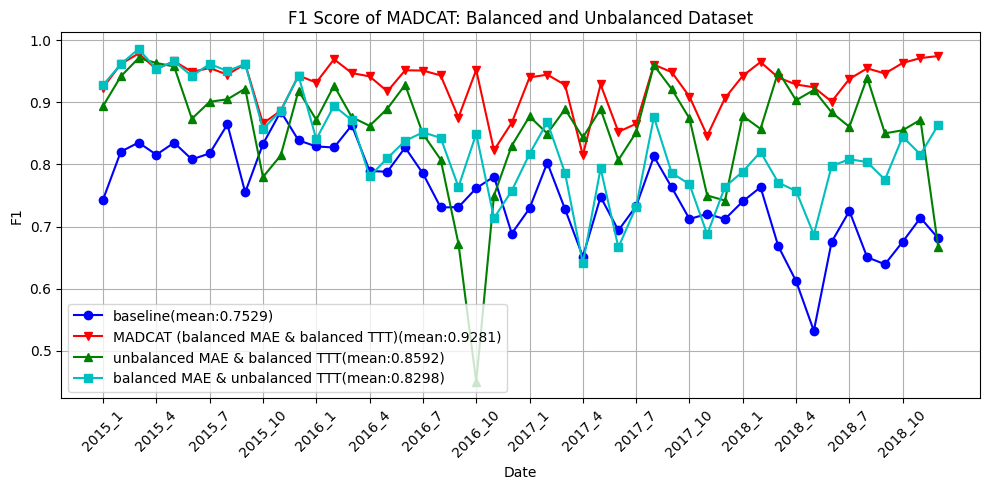}
    \end{subfigure}
    \begin{subfigure}[h]{\linewidth}
        \centering
        \includegraphics[width=\linewidth]{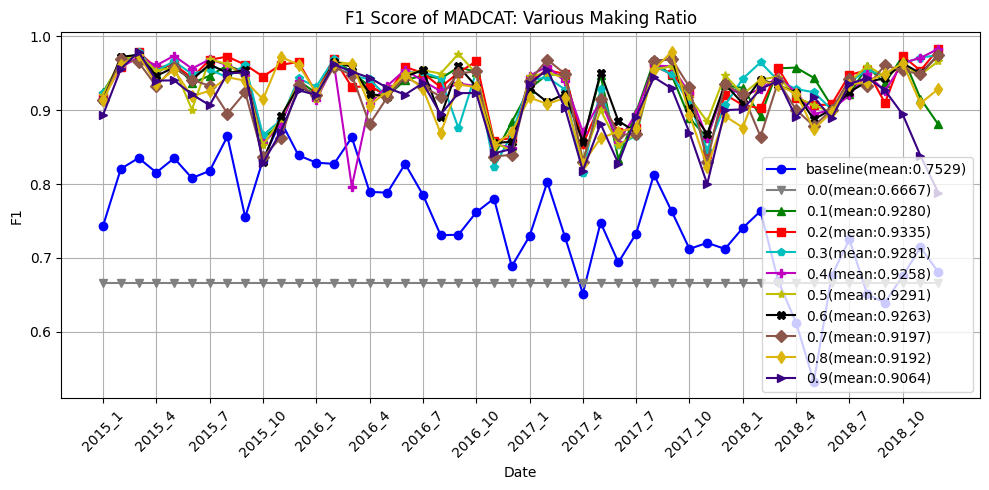}
    \end{subfigure}
    \vspace{-0.9em}
    \caption{\textbf{Impact of database balancing (upper) and masking ratios (lower) on MADCAT performance.}}
    \label{fig:madcat_ablation}
    \vspace{-1.6em}
\end{figure}

\topic{Dataset Balancing.}
We first evaluate the importance of dataset balancing. 
We test the individual impact of balancing during initial training and test-time training. 
For this analysis, we use the original unbalanced dataset, which has a benign-to-malicious sample ratio of 9:1. All other aspects of the experimental setup remain unchanged. The upper plot in Figure~\ref{fig:madcat_ablation} shows the results of our analysis. We first observe that even when trained on an unbalanced dataset, MADCAT remains effective in addressing concept drift in malware detection. In both scenarios, MADCAT consistently outperforms the baseline on average. Among the two, balancing the dataset during initial training results in more stable performance over time.

\topic{Masking Ratio for MAE Training.}
The lower plot in Figure~\ref{fig:madcat_ablation} shows MADCAT’s performance as the masking ratio for MAE training is varied from 0.0--0.9.
We used the balanced dataset for all masking ratio setups.
We first observe that without masking (ratio is 0.0), MAE fails to learn meaningful representations through the reconstruction process. In contrast, when masking is applied (ratios from 0.1--0.9), MADCAT consistently outperforms the baseline. Among these, masking ratios in the range of 0.1--0.6 yield the best results, achieving detection accuracy above 92\%.

\section{Conclusion}
\label{sec:conclusion}

This work presents MADCAT, 
a novel approach that integrates self-supervised learning with test-time adaptation 
to maintain robust malware detection performance under concept drift. 
We demonstrate that MADCAT performs effectively in continuous Android malware detection.
MADCAT requires only a small, balanced subset of data
and does not rely on human-annotated labels for test-time adaptation.

Our approach is promising and opens up new avenues for future research:
A more comprehensive evaluation of MADCAT---%
across diverse malware types (e.g., Windows PE files),
various detection approaches, and alternative self-supervision techniques---%
will deepen our understanding of its effectiveness and generalizability.

\newpage
\section*{Acknowledgment}

E.R and S.H are partially supported by the Google Faculty Research Award 2023.
Y.K is supported by the U.S. Intelligence Community Postdoctoral Fellowship.
The findings and conclusions in this work are those of the author(s) 
and do not necessarily represent the views of the funding agency.

{
    \bibliographystyle{icml2025}    
    \bibliography{bib/thiswork}
}

\newpage
\appendix
\onecolumn

\section{Experimental Setup in Detail}
\label{appendix:detailed-setup}

\topic{Environment.}
All experiments were conducted using Python 3.8.5 and PyTorch 1.11.0 with CUDA 11.3 on a Rocky Linux 9.5. environment.
We run our experiments on an internal cluster with an Intel(R) Xeon(R) Gold 6248R CPUs running at 3GHz with 48 cores and NVIDIA A40 GPUs with 48GB of VRAM.
Our machine has 768GB of DDR4 RAM operating at 2933 MT/s.

\topic{Dataset.}
We used the APIGraph~\cite{chen2023continuous} dataset throughout our evaluation.
Each sample is represented by a binary feature vector of 1,159 dimensions, with 0 or 1 indicating the presence or absence of each feature.
The dataset is labeled as either benign (0) or malicious (1).
Table~\ref{tab:dataset_size_per_year} shows the yearly distribution of samples.

\begin{table}[ht]
\centering
\caption{\textbf{Yearly distribution of the APIGraph dataset.} We use the same dataset as in~\citet{chen2023continuous}.}
\label{tab:dataset_size_per_year}
\begin{tabular}{@{}crrr@{}}
\toprule
Year & \multicolumn{1}{c}{Malicious} & \multicolumn{1}{c}{Benign} & \multicolumn{1}{c}{Total} \\ \midrule \midrule
2012 & 3,061 & 27,472 & 30,533 \\
2013 & 4,854 & 43,714 & 48,568 \\
2014 & 5,809 & 52,676 & 58,485 \\
2015 & 5,508 & 51,944 & 57,452 \\
2016 & 5,324 & 50,712 & 56,036 \\
2017 & 2,465 & 24,847 & 27,312 \\
2018 & 3,783 & 38,146 & 41,929 \\ \bottomrule
\end{tabular}
\end{table}

\section{Additional Evaluation Restuls}
\label{appendix:benign_mal_plot}

\subsection{MADCAT Accuracy per Class Labels}
\label{appendix:madcat_benign_mal}

\begin{figure}[h]
    \centering
    \includegraphics[width=0.48\linewidth]{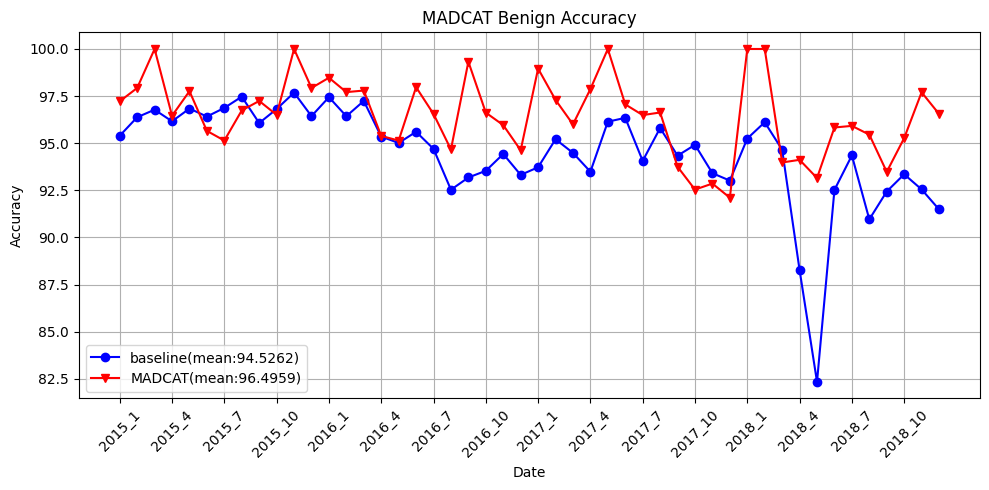}
    \includegraphics[width=0.48\linewidth]{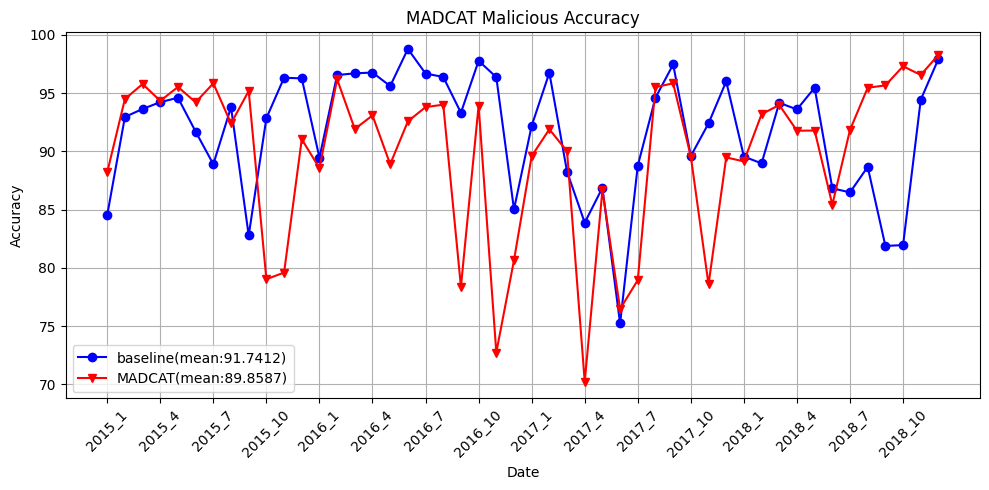}
    \caption{\textbf{MADCAT performance with baseline.} Benign accuracy consistently outperforms the baseline; malicious accuracy is sometimes lower but mostly comparable.}
    \label{fig:madcat_benign_mal}
\end{figure}

Figure~\ref{fig:madcat_benign_mal} shows the class-wise accuracy of MADCAT on benign and malicious samples.
Throughout the evaluation period, the accuracy on benign samples consistently exceeds that of the baseline.
For malicious samples, MADCAT occasionally shows slightly lower accuracy than the baseline; however, its performance remains mostly comparable across all datasets.

\subsection{MADCAT Accuracy with Pseudo-Label}
\label{appendix:madcat_pseudo_benign_mal}

\begin{figure}[ht]
    \centering
    \includegraphics[width=0.48\linewidth]{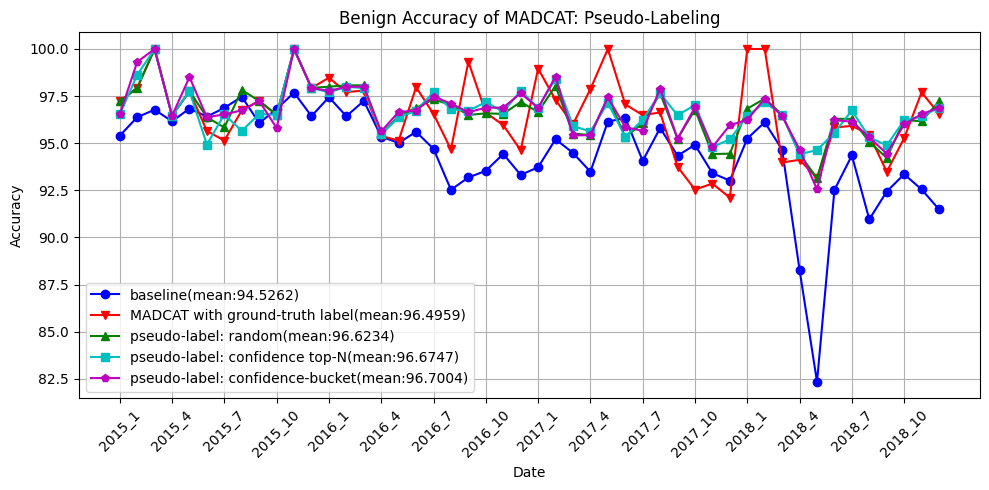}
    \includegraphics[width=0.48\linewidth]{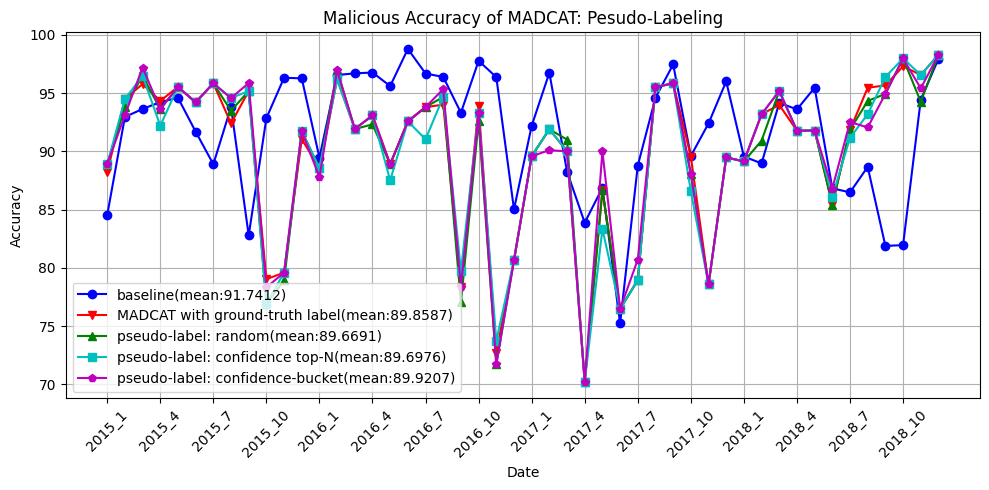}
    \caption{\textbf{Performance of MADCAT with different pseudo-labeling strategies.} All methods show similar or higher benign accuracy than the baseline; malicious accuracy is slightly lower but comparable.}
    \label{fig:madcat_pseudo_label_benign_mal}
\end{figure}

Figure~\ref{fig:madcat_pseudo_label_benign_mal} shows the class-wise accuracy of MADCAT when using pseudo-labeling for dataset balancing.
Similar to the results with ground-truth-based balancing, pseudo-labeling improves the benign accuracy and yields comparable or slightly lower accuracy on malicious samples.
Overall, pseudo-labeled MADCAT achieves results similar to those using ground-truth labels, indicating its robustness even under label-free conditions.

\end{document}